\documentstyle[epsfig]{l-aa}
\def\h0{{\rm H_0}}

\def\oiii{[O\kern.2em{\sc iii}] }
\def\oiiinb{[O\kern.2em{\sc iii}]}
\def\oiiif{[O\kern.2em{\sc iii}] \kern.2em{3p2-3p1}}
\def\oiiie{[O\kern.2em{\sc iii}] \kern.2em{3p1-3p0}}
\def\oiv{[O\kern.2em{\sc iv}] } 
\def\oivt{[O\kern.2em{\sc iv}] \kern.2em{2p$\frac{3}{2}$-2p$\frac{1}{2}$} } 
\def\oi{[O\kern.2em{\sc i}] } 
\def\ois{[O\kern.2em{\sc i}] \kern.2em{3p1-3p2}} 
\def\oio{[O\kern.2em{\sc i}] \kern.2em{3p0-3p1}} 

\def\cii{[C\kern.2em{\sc ii}] }
\def\ciio{[C\kern.2em{\sc ii}] \kern.2em{2p$\frac{3}{2}$-2p$\frac{1}{2}$}}

\def\ariii{[Ar\kern.2em{\sc iii}] } 
\def\ariiin{[Ar\kern.2em{\sc iii}] \kern.2em{3p1-3p2}} 
\def\arii{[Ar\kern.2em{\sc ii}] } 
\def\ariis{[Ar\kern.2em{\sc ii}] \kern.2em{2p$\frac{1}{2}$-2p$\frac{3}{2}$} } 

\def\neii{[Ne\kern.2em{\sc ii}] }
\def\neiit{[Ne\kern.2em{\sc ii}] \kern.2em{2p$\frac{1}{2}$-2p$\frac{3}{2}$} }
\def\neiii{[Ne\kern.2em{\sc iii}] } 
\def\neiiif{[Ne\kern.2em{\sc iii}] \kern.2em{3p1-3p0}} 

\def\niii{[N\kern.2em{\sc iii}] } 
\def\niiif{[N\kern.2em{\sc iii}] \kern.2em{2p$\frac{3}{2}$-2p$\frac{1}{2}$} }
\def\nii{[N\kern.2em{\sc ii}] } 
\def\niif{[N\kern.2em{\sc ii}] \kern.2em{3p2-3p1}}

\def\siii{[S\kern.2em{\sc iii}] }  
\def\siiie{[S\kern.2em{\sc iii}] \kern.2em{3p2-3p1}}
\def\siiit{[S\kern.2em{\sc iii}] \kern.2em{3p1-3p0}}
\def\siv{[S\kern.2em{\sc iv}] } 
\def\sivt{[S\kern.2em{\sc iv}] \kern.2em{2p$\frac{3}{2}$-2p$\frac{1}{2}$} } 

\begin{document}

%

\thesaurus{03(11.09.1 Centaurus A = NGC 5128; 13.09.1;11.09.4; 11.19.3,11.01.2)}
\title{ISO-LWS spectroscopy of Centaurus A: extended star formation}

\author{ S. J. Unger \inst{1}  
\and P. E. Clegg \inst{1}
\and G. J. Stacey \inst{2}
\and P. Cox \inst{3}
\and J. Fischer \inst{4} 
\and M. Greenhouse \inst{5} 
\and S. D. Lord \inst{6}
\and M. L. Luhman \inst{4} 
\and S. Satyapal \inst{5} 
\and H. A. Smith  \inst{7}
\and L. Spinoglio \inst{8}
\and M. Wolfire \inst{9} }

\offprints{s.j.unger@qmw.ac.uk}

\institute{Physics Dept., Queen Mary \& Westfield College, University of 
London, London E1 4NS, U.K.
\and Cornell University, Ithaca, NY, USA
\and Institut d'Astrophysique Spatiale, Orsay, France
\and Naval Research Laboratory, Washington, USA 
\and NASA Goddard, Greenbelt, USA
\and IPAC, California Institute of Technology, Pasadena, USA
\and Harvard-Smithsonian Center for Astrophysics, Cambridge, MA, USA
\and Istituto di Fisica dello Spazio Interplanetario-CNR, Roma, Italy
\and University of Maryland, College Park, MD, USA }

\date{Received ;Accepted}

\maketitle

\begin{abstract}
We present the first full FIR spectrum of Centaurus A (NGC 5128)
from 43 - 196.7 $\mu$m. The data was obtained with the ISO Long Wavelength
Spectrometer (LWS). We conclude that the FIR emission in a 
70~\arcsec~beam centred on the nucleus is dominated by star formation
rather than AGN activity. The flux in the 
far-infrared  lines is $\sim$ 1 \% of the total FIR: 
the \cii line flux is $\sim$ 0.4 \% FIR and the \oi line is 
$\sim$ 0.2 \%, with the remainder arising from \oiiinb, \nii and
\niii lines. These are typical values for starburst galaxies.

The ratio of the \niii / \nii line intensities from the HII regions in
the dust lane corresponds to an effective temperature, 
T$_{\mathrm{eff}}$ $\sim$ $35\,500$ K, implying that the tip of the 
main sequence is headed by O8.5 stars and that the starburst is 
$\sim$ 6 $\times 10^6$ years old. This suggests that the galaxy
underwent either a recent merger or a 
merger which triggered a series of bursts. The N/O abundance ratio is
consistent with the range of $\sim$ 0.2 - 0.3 found for Galactic HII regions. 

We estimate that $<$ 5 \% of the observed \cii arises in the cold
neutral medium (CNM) and
that $\sim$ 10 \% arises in the warm ionized medium (WIM). The main
contributors to the 
\cii emission are the PDRs, which
are located throughout the dust lane and in regions beyond where the bulk of
the molecular material lies. On scales of $\sim$ 1 kpc the average physical 
properties of the PDRs are modelled with a gas density, 
n $\sim$ $10^3$ cm$^{-3}$, an incident far-UV field, G $\sim$ $10^2$
times the local Galactic field, and a 
gas temperature of $\sim$ 250 K. 
\end{abstract}

\section{Introduction}

Centaurus A (NGC 5128) is the nearest (d = 3.5 Mpc; 1 \arcsec $\sim$17~pc, Hui 
et al. 1993) example of a giant elliptical galaxy associated with a powerful 
radio source. The large-scale radio morphology consists of twin radio
lobes separated by
$\sim$ 5 degrees on the sky. The compact ($\sim$ milliarcsecond) radio 
nucleus is variable and has a strong jet extending 
$\sim$ 4 arcminutes towards the 
northeast lobe. The spectacular optical appearance is that of a 
giant elliptical galaxy that appears enveloped in a nearly edge on, warped 
dust lane. There is also a series of faint optical shells.
The stellar population in the dominant elliptical structure is old, whilst 
that of the twisted dust lane is young, sporadically punctuated by HII 
regions, 
dust and gas (Graham 1979). The overall structure of Cen A resembles
that of a recent ($< 4 \times 10^8$ years, Tubbs 1980) merger, 
between a spiral and a large elliptical galaxy. The dust lane is the source 
of most (90 \%) of the far-infrared luminosity (L$_{\mathrm{FIR}} \sim 3 
\times 10^{9}$ L$_{\odot}$) and is thought to be re-radiated starlight 
from young stars in the dusty disk (Joy et al. 1988).

In Sect. 2 we describe the observations and data analysis. Sect. 3 looks at
the general FIR properties and proceeds to model the
HII regions and the PDRs in the dust lane. Sect. 4 summarises the results
and presents our conclusions.

\section{Observations}

Cen A was observed with the LWS grating 
($R=\lambda/\Delta\lambda \sim 200$) as part of the LWS consortium's
guaranteed time extragalactic programme. A full grating observation 
(43 - 196.7 $\mu$m) was taken of the nucleus at the centre of the dust
lane and a series of line observations were taken at two positions in
the SE and NW regions of the dust lane. 
A short \cii~157 $\mu$m line observation was taken off-source at
position \#4 (see Table 1) to estimate the Galactic emission near the 
source. Position \#1 was intended to provide a deeper integration 
coincident with position \#2, but was accidently offset.

A series of half-second integration ramps were taken at each grating
position with four samples per resolution element ($\Delta\lambda =
0.29~\mu$m $\lambda\lambda 43 - 93~\mu$m and $\Delta\lambda = 0.6~\mu$m 
$\lambda\lambda 84 - 196~\mu$m). The total integration time per
resolution element and per pointing were:
position \#1 88~s for the \oiii 52~$\mu$m and 34~s for the 
\niii 57~$\mu$m; 
position \#2 (the centre), 30~s for the range  43--196 $\mu$m; 
positions NW and SE  (2 point raster
map) 22~s for the the \oi 63~$\mu$m, 14~s for the \oiii 88~$\mu$m, 
12~s for the \nii 122~$\mu$m, 28~s for the \oi 145~$\mu$m and 12~s for
the \cii 158~$\mu$m; 
position \#4 12~s for the \cii 158~$\mu$m.

The data were processed with RAL pipeline 7 and analysed using the LIA
and ISAP packages. 
The LWS flux calibration and relative spectral response function (RSRF)
were derived from
observations of Uranus (Swinyard et al. 1998). The full grating spectrum at
the centre enabled us to estimate the relative flux uncertainty
between individual detectors arising from 
uncertainties in the relative responsivity and the 
dark-current subtraction. 
The offsets between the detectors (excluding detector
SW1) was $\leq 10$ \%. The \oiii 88 $\mu$m line on detectors SW5 and LW1 
had a 15 \%
systematic uncertainty and the \cii line on detectors LW3 and LW4 had a 
10 \% systematic uncertainty. We therefore adopt a relative flux uncertainty 
of $\sim$ 15\%. Because we only took spectra of individual lines at the
NW and SE positions there is no corresponding overlap in wavelength
coverage at these positions. One indicator of relative flux uncertainty 
is a discrete step down in flux, of $\sim$ 25 \%, 
at $\sim$ 125~$\mu$m at 
the SE position. The relative flux uncertainty is assumed to be
$\le$ 25 \%  at these positions.

The absolute flux calibration w.r.t. Uranus for point like objects 
observed on axis is better than 15 \% (Swinyard et al. 1998). However, extended
sources give rise either to channel fringes or to a
spectrum that is not a smooth function of wavelength. This is still a
calibration issue. For example, in 
Fig. 2, detectors SW5, LW1, LW2 have slopes that differ from those of
their neighbours in the overlap region. This may account for the
continuum shape, which is
discussed in Sect. 3.1.
The LWS beam profile is known to be asymmetric and is still under
investigation. We therefore adopt a value for the FWHM of 70~\arcsec~at
all wavelengths, believing that a more sophisticated treatment would not
significantly affect our conclusions. We also note that there is good 
cross calibration between the ISO-LWS results and the Far-infrared Imaging
Fabry-Perot Interferometer (FIFI)  (Madden et al. 1995); 
the \cii peak fluxes agree to within $\sim$ 10 \%.

\begin{table}
\begin{center}
\caption{ Observation Log}
\vspace*{0.3cm}
\begin{tabular}{l|r|r|l}
{Position}&{Offset in RA}&{Offset in Dec}&{Date/AOT}\\
&arcsec&arcsec&\\
\hline
&&&\\
\#1&+ 29&- 12 &1996 Aug 23 L02\\
\#2 Centre&0&0&1997 Aug 11 L01\\ 
\#3 (map)&& &1997 Feb 12 L02\\
NW&- 53&+ 27&\\ 
SE&+ 110&- 49&\\
\#4 (off)&- 2 & + 600&1997 Feb 12 L02\\
\end{tabular}

Offsets w.r.t. 13h 25m 27.6s -43d 01 \arcmin 08.6 \arcsec J2000\\
\end{center}
\end{table}
\normalsize

\begin{figure}[h]
  \begin{center}
    \leavevmode
 \centerline{\epsfig{file=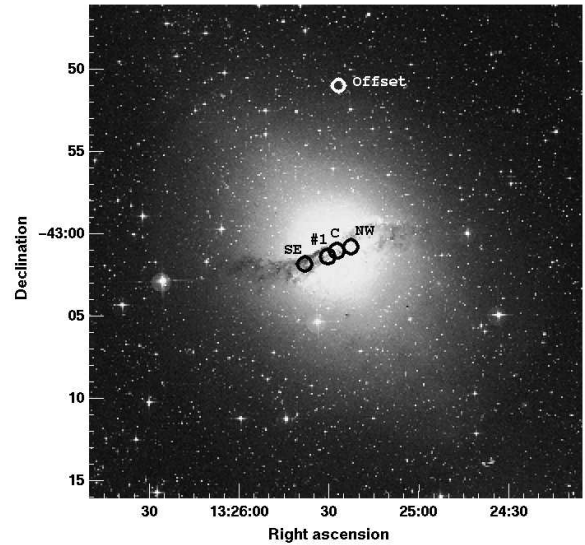,width=8.0cm,angle=0}}
  \end{center}
 \caption{\rm Cen A digital sky survey image overlaid with the LWS beam
  positions}
  \label{fig:sample1}
\end{figure}

\section{Results and discussion}

\subsection{\rm General FIR properties}

The far-infrared continuum at each position is modelled with a
single-temperature blackbody spectrum of the form F$_{\lambda}$ $\alpha$ 
$\Omega$  B$_{\lambda}$(T)(1-e$^{-\tau_{\mathrm{dust}}}$), where the
solid angle,
$\Omega$, is constrained to equal the LWS beam, B$_{\lambda}$(T) is the 
Planck function at temperature T and $\tau_{\mathrm{dust}}$ $\alpha$ 
$\lambda^{-1.5}$. The result for the central position is shown as the
dashed curve in Fig. 2. Although the observed continuum is not a
simple function of wavelength and the single temperature blackbody is
not an especially good fit, particularly at wavelengths $> 100 \mu$m, a
better calibration of straylight and the beam profile is required for
anything more sophisticated. The best FIR temperature at each position
is $\sim$ 30 K.

\begin{figure}[h]
  \begin{center}
    \leavevmode
 \centerline{\epsfig{file=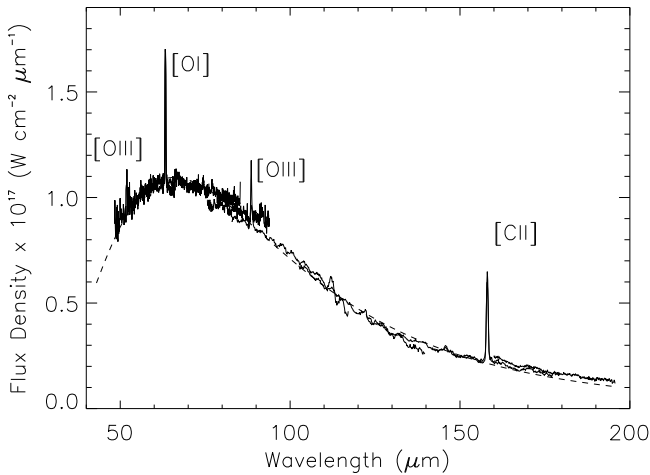,width=7.cm,angle=90}}
  \end{center}
 \caption{\rm LWS spectrum of the central region of the dust lane
(dashed line is the blackbody fit)}
  \label{fig:saple1}
\end{figure}

The luminosities quoted here are derived from the line fluxes listed in
Table 2. At the central position, the total luminosity in all of the 
far-infrared  lines 
is 2.6 $\times 10^7$ L$_{\odot}$, which is $\sim$ 1 \% of the total FIR 
luminosity (L$_{43-197 \mu m} = 3.2 \times 10^9$ L$_{\odot}$). 
The \cii luminosity is 
1.1 $\times 10^7$ L$_{\odot}$ (0.4 \% FIR) and the \oi luminosity is 
7.5 $\times 10^6$ L$_{\odot}$ (0.2 \% FIR). 
Because full spectra are not available at the NW and SE positions we 
estimate the FIR continuum luminosity by integrating under the 
single-temperature blackbody fit
to the data.
At the NW position the total FIR luminosity,
L$_{43-197 \mu m} = 2.2 \times 10^9$ L$_{\odot}$. The \cii luminosity is 
9.3 $\times 10^6$ L$_{\odot}$ (0.4 \% FIR) and the \oi luminosity is 
3.4 $\times 10^6$ L$_{\odot}$ (0.2 \% FIR).
At the SE  position the total FIR luminosity,
L$_{43-197 \mu m} = 8.7 \times 10^8$ L$_{\odot}$. The \cii luminosity is 
3.5 $\times 10^6$ L$_{\odot}$ (0.4 \% FIR) and the \oi luminosity is 
2.0 $\times 10^6$ L$_{\odot}$ (0.2 \% FIR). These are typical values
for starburst galaxies (c.f. Lord et al. 1996).

\begin{table*}
\begin{center}
\caption{Line Fluxes}
\vspace*{0.3cm}
\begin{tabular}{l|l|l|l|l|l}
{Line}&{$\lambda_{\rm{rest}}$}&{\#1}&{Centre}&{NW}&{SE}\\
&$\mu$m&Flux&Flux&Flux&Flux\\
\hline
&&&&&\\
\oiiif&51.815&0.31&0.72&-&-\\
\niiif&57.3170&0.16&0.24&-&-\\
\ois&63.184&-&1.96&0.90&0.51\\
\oiiie&88.356&-&0.70&0.6&0.2\\
\niif&121.898&-&0.15&0.15&$\le$0.15\\
\oio&145.525&-&0.11&0.08&$\le$0.03\\
\ciio&157.741&-&2.91&2.43&0.92\\
\end{tabular}

\normalsize
Flux $ \times10^{-18}$ W cm$^{-2}$\\
- wavelength range not covered\\
upper limits are 3 x rms residuals from a fit to the continuum \\
\cii~157~$\mu$m flux at \#4 (off) is 0.04 \\
\end{center}
\end{table*}
\normalsize

\subsection{\rm Ionized gas lines}

Photons of energy 35.12, 29.60 and 14.53 eV are required to form 
O++, N++ and N+, 
respectively, so that the observed \oiii, \niii and \nii 
emission must originate in or around HII regions. The \oiii line ratio
is a sensitive 
function of density in the range $\sim 30 - 10^4$ cm$^{-3}$.
For the central position this ratio is $\sim$ 0.9, corresponding
to an electron density, n$_{\mathrm{e}} \sim$ 100 cm$^{-3}$ (Rubin et
al. 1994). 
The \oiii lines indicate a higher electron density, 
n$_{\mathrm{e}} \sim$ 250 cm$^{-3}$, for the starburst nuclei of M82 
(Colbert et 
al. 1999) and M83 (Stacey et al. 1999). In contrast, the \nii 
205 $\mu$m / 122 $\mu$m line intensity ratio for the Galaxy gives
an average electron density, of only $\sim$ 3 cm$^{-3}$ 
(Petuchowski \& Bennett 1993).
The {\it thermal pressure} of the ionized material in the Cen A dust 
lane is therefore closer to that of starburst galaxies than to that of the 
Milky Way.

Since the \niii 57 $\mu$m  and the \nii 122 $\mu$m lines arise from 
different ionization states of the same element, the line intensity 
ratio is sensitive to the hardness of the interstellar UV field and
therefore to the spectral type of the hottest main sequence star. 
For the central 
position \niii / \nii $\sim$ 1.6. This is larger 
than the value of $\sim$ 0.9 for M83 (Stacey et al. 1999) but smaller
than the value of $\sim$ 2.1 for M82 (Colbert et al. 1999). 
Assuming that the region 
is ionization bounded, with an electron density, n$_e$ $\sim$ 100 cm$^{-3}$
the \niii / \nii line intensity 
ratio for Cen A corresponds to an abundance ratio N++/N+ of $\sim$ 0.3; this 
corresponds to an effective temperature, T$_{\mathrm{eff}} \sim 35\,500$ K 
(Rubin et al. 1994). Applying the same 
corrections to M82 and M83 with n$_e$ $\sim$ 250 cm$^{-3}$ implies an
effective temperature, T$_{\mathrm{eff}} \sim 34\,500$ K for M83 and 
T$_{\mathrm{eff}} \sim$ 35\,500 K for M82. If the effective temperature
in Cen A corresponds to the tip of the main sequence formed in a single
starburst, we are observing  O8.5 stars and the burst is 
$\sim$ 6 $\times 10^6$ years old.
If the burst was triggered by the spiral-elliptical galaxy merger 
then its occurance was very recent. Alternatively, the merger triggered 
a series of bursts of star formation, of which we are witnessing the most
recent.

The N++ and O++ coexist in roughly the same ionization zones, and the
\oiii 52 $\mu$m and \niii 57 $\mu$m lines have roughly the same critical 
density. As a result the ratio of these lines is an indicator, to within
$\sim 50$~\%, of the N++/O++ 
abundance ratio, which itself, is nearly equal to the N/O ratio in the hard UV 
field environments we are seeing here (Rubin et al. 1988). 
The line ratio we observe at 
the  centre of the dust lane is $\sim$ 0.3 - the same as found in the
nucleus of M82 (Colbert et al. 1999), but much smaller than that found for 
the nucleus of M83 ($\sim$ 0.67 Stacey et al. 1999). 
 
A more precise determination of the abundance ratio requires the 
observed line ratio to be divided by the volume emissivity ratio. The latter 
ratio is dependent on the electron 
density because the two lines have slightly different critical densities.
Using our value for the electron density $\sim$ 100 cm$^{-3}$ and 
Fig. 3 of Lester et al. (1987) we estimate that the N/O abundance
ratio to be $\sim$ 0.2 in Cen A. This value is consistent with the range of 
$\sim$ 0.2 - 0.3 found for Galactic HII regions (Rubin et al. 1988). The
nitrogen to oxygen abundance ratio is a 
measure of the chemical evolution and we expect it to increase with time 
(cf. the solar value of $\sim$ 0.12).

\subsection{\rm Neutral gas lines}

Carbon has a low ionization potential (11.4 eV), which is less than that
of hydrogen. \cii 157~$\mu$m line emission
is therefore observed from both neutral and ionized hydrogen clouds. We
model the \cii line emission with three components:
Photodissociation regions (PDRs) on the surfaces of UV exposed
molecular clouds; cold (T $\sim$ 100 K) HI clouds (i.e. the cold neutral 
medium (CNM) Kulkarni \& Heiles 1987, Wolfire et al. 1995); and
diffuse HII regions (i.e. the warm ionized medium (WIM) Heiles 1994).

\subsubsection{\rm HI clouds}

It can be shown that the intensity in the \cii line emitted from gas clouds 
with density, n(H) and temperature (T) is given by (c.f. Madden et al. 1993)

\begin{equation}
\mathrm
I_{c^+} = 2.35 \times 10^{-21} \big[ \frac{2exp(\frac{-91.3}{T})}{1 + 2exp(\frac{-91.3}{T}) + \frac{n_{crit}}{n_H}} \big] X_{c^+}N(HI)
\end{equation}

where the critical density for collisions with H, n$_{\mathrm{crit}} 
\sim 3000$ cm$^{-3}$ (Launay \& Roueff 1977) and the fractional C$^+$  
relative to hydrogen is $\mathrm{X_{c^+} \sim  X_{c} = 1.4 \times
10^{-4}} $ (Sofia et al. 1997).

N(HI) is estimated from the HI 21cm map of Van Gorkom et al. (1990) to be
18.8 $\times 10^{20}$ atom cm$^{-2}$ at the SE position.  The central and
NW positions are difficult to estimate due to HI absorption against the
nuclear continuum. There may be a central hole in the HI and the column
density is certainly not higher than the peak observed in the SE region 
of the dust lane (Van Gorkom et al. 1990) 

Assuming typical Galactic values for the temperature, T
$\sim$ 80 K, and hydrogen density, n $\sim$ 30 cm$^{-3}$, results in an
estimated \cii flux of $3.5 \times 10^{-20}$ W cm$^{-2}$ in a
70~\arcsec~LWS beam at the SE position. This corresponds to 4 \%, 1 \%
and 1 \%
of the observed \cii  flux at positions SE, Centre and NW respectively.
The peak HI emission line flux corresponds to 
$1.9 \times 10^{-19}$ W cm$^{-2}$ which is only 6~\% and 8~\% of the 
\cii flux at the centre and NW positions respectively. We conclude that
there is very little \cii emission in our beams from HI clouds.

\subsubsection{\rm Diffuse HII regions}

Ionized carbon can be found in both neutral gas and ionized gas clouds, 
and is an important coolant for each. We detected \oiii 88 $\mu$m in all 3 
beam positions so there is an ionized gas component in each
beam. Using the constant density HII region model of Rubin (1985) with the 
Kurucz abundances, 10$^{49}$ ionizing photons per second and our derived
density, n$_e \sim$ 100 cm$^{-3}$ and effective 
temperature, T$_{\mathrm{eff}} \sim$ 35\,500 K we can estimate the 
\cii emission
from the HII regions. Applying the model \oiii 88
$\mu$m / \cii 158 $\mu$m line ratio of 0.35 to the observed \oiii 88
$\mu$m line flux at each position results in $\sim$ 10 \% contribution
to the observed \cii line flux in each beam. Scaling the model fluxes to
the distance of Cen A gives $\sim$ 3000 HII regions in the central and
NW regions and $\sim$ 1000 HII regions in the SE region. 

The estimate above assumes that the observed lines have the same filling
factor in the large LWS beam. If, alternatively, we were to assume that
the ionized component was instead dominated by a contribution 
from an extended low density warm ionized medium (ELDWIM) with 
n$_e \sim$ 3 cm$^{-3}$, then the \cii flux can be
estimated from the ratio of the  \cii/\nii lines to be $\sim$ 18 \% at
the central position. The observations of the \nii 121.9~$\mu$m line at
the NW and SE positions (with lower signal to noise)
indicate a similar fractional component (21 \% and $\le 56$ \%, respectively).

We have estimated the density in the HII regions in the centre of 
Cen A to be $\sim$ 100 cm$^{-3}$ with an effective temperature, 
T$_{\mathrm{eff}} \sim 35\,500$ K. Based on the HII region models of
Rubin (1985) we estimate that $\sim$ 10 \% of the observed \cii arises
in the WIM. 

\begin{figure}[h]
  \begin{center}
    \leavevmode
 \centerline{\epsfig{file=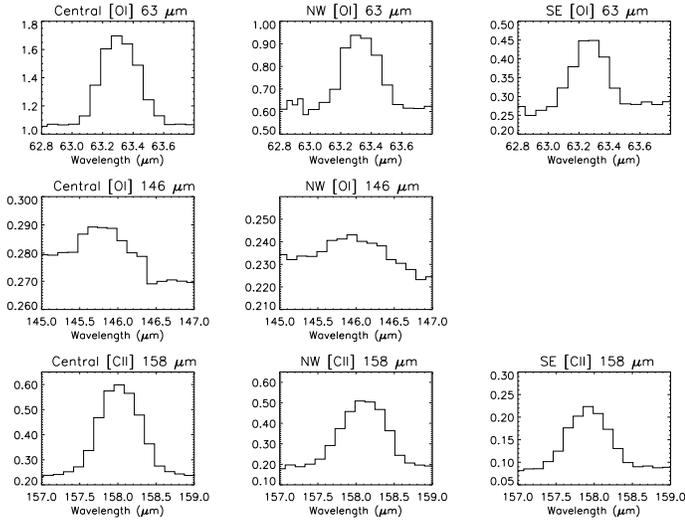,width=7.0cm,angle=90}}
  \end{center}
 \caption{\rm PDR Lines (Flux Density $\times 10^{17}$ W cm$^{-2}$
  $\mu$m$^{-1}$). Note the \cii lines are the total observed flux density
  i.e. PDR + CNM + WIM}
  \label{fig:saple1}
\end{figure}

\subsubsection{\rm PDRs}

Far-UV photons (6 eV $<$ h$\nu \leq$ 13.6 eV) from either O/B stars or
an AGN will photo-dissociate H$_2$ and CO molecules and photo-ionize
elements with ionization potentials less than the Lyman limit
(e.g. C$^+$ ionization potential = 11.26 eV). The gas heating in these 
photodissociation regions (PDRs) is dominated by electrons ejected from 
grains due to the
photoelectric effect. Gas cooling is dominated by the emission of
\oi~63~$\mu$m and \cii~158~$\mu$m emission.
Observations of these lines,
the \oi~146~$\mu$m and CO~(J=1-0) 2.6 mm lines and the FIR continuum can be
used to model the average physical properties of the neutral
interstellar medium (Wolfire et al. 1990). Kaufman et al. (1999) have
computed PDR models over a wide range of physical conditions. The new
code accounts for gas heating by
small grains/PAHs and large molecules, and uses a lower,
gas phase carbon abundance  (X$_{\mathrm{C}}$ = 1.4x10$^{-4}$, Sofia et
al. 1997)
and oxygen abundance (X$_{\mathrm{O}}$ = 3.0x10$^{-4}$, Meyer et al. 1998).
The \oi 63 $\mu$m / \cii 158 $\mu$m line ratio and either the \oi 146 $\mu$m
/ \oi 63 $\mu$m line ratio or the (\oi 63 $\mu$m + \cii 158 $\mu$m) /
FIR continuum can be used as PDR diagnostics to determine the 
average gas density (i.e the proton density,
n cm$^{-3}$), the average incident far-UV flux (in units of the Milky Way
flux, G$_o$ = 1.6 $\times 10^{-3}$~erg cm$^{-2}$ s$^{-1}$) and
the gas temperature. 

We assume that the measured \cii flux at each position should have 
$\sim$ 10 \% subtracted, due to the HI and WIM components, before it is used 
to model the PDRs (if, alternatively, a 20 \% ELDWIM contribution is 
subtracted it would not significantly affect the PDR parameters derived 
below). The PDR lines are plotted in Fig. 3 and the line intensity
ratios are given in Table 3.

\begin{table}
\begin{center}
\caption{PDR diagnostic line intensity ratios}
\vspace*{0.3cm}
\begin{tabular}{l|l|l|l}
{Line Ratio}&{Centre}&{NW}&{SE}\\
\hline
&&&\\
\oi 63~$\mu$m  / \cii 158~$\mu$m &0.7&0.4&0.6\\
\oi 146~$\mu$m  / \oi 63~$\mu$m &0.06&0.09&$\le$ 0.06\\
(\oi 63~$\mu$m  + \cii 158~$\mu$m )/FIR&0.006&0.005&0.006\\
\end{tabular}
\normalsize
\end{center}
\end{table}
\normalsize


The results for the three regions are
consistent with each other, having a gas
density, n $\sim$ 10$^3$ cm$^{-3}$, and an incident far-UV field, 
G $\sim$ 10$^2$. 

At the NW position, only the combination of the \oi 63 $\mu$m / 
\cii 158 $\mu$m ratio and the
(\oi 63 $\mu$m + \cii 158 $\mu$m) /FIR continuum ratio gives a 
meaningful solution for G and n. The \oi 146 $\mu$m line is clearly detected
but with a very rippled baseline due to channel fringes. 
The observed \oi 146 $\mu$m line
flux would need to be reduced by $\sim$ 60 \%  in order
to obtain a consistent result with the \oi 146 $\mu$m / \oi 63 $\mu$m
line ratio predicted by the PDR model. 


The LWS results for the nucleus confirm those previously derived from
IR, submm and CO observations. 
The consistent set of derived PDR conditions for all three positions 
suggest that the observed FIR emission
in a 70~\arcsec~beam centred on the nucleus is dominated by star formation
and not AGN activity. Joy et al. (1988) mapped Cen A at 50 and 100
$\mu$m on the KAO. They concluded that the extended FIR
emission was from dust grains heated by massive
young stars distributed throughout the dust lane, not the 
compact nucleus. Hawarden et al. (1993)
mapped Cen A at 800 $\mu$m and 450 $\mu$m with a resolution of 
$\sim$10 \arcsec. 
They attribute the large scale 800 $\mu$m emission to thermal emission
from regions of star formation embedded in the dust lane.
They note that the H$_2$ emission within a few arcseconds of the nucleus,
observed by Israel et al. (1990), indicates that significant UV
radiation from the nucleus does not reach large radii in the plane of
the dust lane i.e. the nuclear contribution to exciting the extended gas and 
dust disk is small.

Eckart et al. (1990) and Wild et al. (1997)
mapped Cen A in $^{12}$CO J=1-0, 
$^{12}$CO J=2-1 and  $^{13}$CO J=1-0. All three maps have two peaks
separated by $\sim$ 90 \arcsec centred on the nucleus.
It is interesting to note that our SE position only clips the lowest
contours of the CO (1-0) and CO (2-1) maps of Wild et
al. (1997). In spite of this the derived PDR parameters are consistent with
those encompassing the bulk of the molecular emission. There must be
extended low level CO (1-0) emission beyond the sensitivity limits of
the Wild et al. (1997) maps. The lowest contour is 17.5 K kms$^{-1}$,
corresponding to M$_{\mathrm{H}_2}$ $\sim$ 10$^{8}$ M$_{\odot}$ if the material
filled the LWS beam.

\section{Summary and conclusions}

We present the first full FIR spectrum from 43 - 196.7 $\mu$m of Cen A.
We detect seven fine structure lines (see Table 2), the strongest being 
those generated in PDRs.
At the central position, the total flux in the far-infrared  lines 
is $\sim$ 1 \% of the total FIR luminosity 
(L$_{43-197 \mu m} = 3.2 \times 10^9$ L$_{\odot}$ for a distance of 3.5 Mpc). 
The \cii line flux is $\sim$0.4 \% FIR and the \oi line flux is 
$\sim$ 0.2 \% FIR. These are typical values for starburst galaxies (Lord 
et al. 1996). The \oiii 52 $\mu$m / \oiii 88 $\mu$m line 
intensity ratio is $\sim$ 0.9, which corresponds to an electron density, 
n$_{\mathrm{e}} \sim$ 100 cm$^{-3}$ (Rubin et al. 1994). 
The {\it thermal pressure} of the ionized medium in the Cen A dust 
lane is closer to that of starburst galaxies (n$_e \sim$ 250 cm$^{-3}$ in 
M82 (Colbert et al. 1999) and M83 (Stacey et al. 1999)) than that of the 
Milky Way (n$_e \sim$ 3 cm$^{-3}$ (Pettuchowski \& Bennett 1993)).

The \niii / \nii line intensity 
ratio is $\sim$ 1.6, giving an abundance ratio N++/N+ $\sim$ 0.3, which 
corresponds to an effective temperature, T$_{\mathrm{eff}} \sim$ 35\,500 K 
(Rubin et al. 1994). Assuming a coeval 
starburst, then the tip of the main sequence is headed by O8.5
stars, and the starburst is $\sim$ 6 $\times 10^6$ 
years old. 
If the burst in Cen A was triggered by the spiral-elliptical galaxy merger 
then its occurance was very recent. Alternatively, the merger triggered 
a series of bursts of star formation and we are witnessing the most recent 
activity. 

We estimate that the N/O abundance ratio is $\sim$ 0.2 in the HII regions in 
Cen A. This value is consistent with the range of 
$\sim$ 0.2 - 0.3 found for Galactic HII regions (Rubin et al. 1988). N/O is a 
measure of the chemical evolution and we expect it to increase with time
(c.f. the solar value of $\sim$ 0.12).

We estimate that $\sim$ 10 \% of the observed \cii arises in the 
WIM. The CNM contributes very little ($< 5$ \%) \cii emission at our
beam positions. The bulk of the emission is from the PDRs.

We derive the average physical conditions for the PDRs in Cen A for
the first time. There is active star formation throughout the dust lane
and in regions beyond the bulk of the molecular material. The FIR emission in 
the 70~\arcsec~LWS beam at the nucleus is dominated by emission from 
star formation rather than AGN activity. On scales of
$\sim$ 1 kpc the average
physical properties of the PDRs are modelled with a gas
density, n $\sim$ 10$^3$ cm$^{-3}$, an incident far-UV field, 
G $\sim$ 10$^2$ and a gas temperature of $\sim$ 250 K.

\section*{Acknowledgements}
Many thanks to the dedicated efforts of the LWS instrument team. The ISO 
Spectral Analysis Package (ISAP) is a joint development by the
LWS and SWS Instrument Teams and Data Centers. Contributing institutes
are CESR, IAS, IPAC, MPE, QMW, RAL and SRON.

\section*{References}

Colbert J.W., Malkan M.A., Clegg P.E., et al., 1999, ApJ 511, 721 \\
Eckart A., Cameron M., Rothermel H., et al., 1990, ApJ 363, 451\\
Graham J., 1979, ApJ 232, 60\\
Hawarden T.G., Sandell G., Matthews H.E., et
al., 1993, MNRAS 260, 844\\
Heiles C., 1994, ApJ 436, 720\\
Hui X., Ford H.C., Ciardillo R., et al., 1993, ApJ 414, 463\\
Israel F.P., van Dishoeck E.F., Baas F. et al.,
1990, A\&A 227, 342\\
Joy M., Lester D.F., Harvey P.M., et al., 1988, ApJ 326, 662\\
Kaufman M.J., Wolfire M.G., Hollenbach D., et al., 1999, ApJ in press\\
Kulkarni S.R., Heiles C., 1987, in Hollenbach, D., Thronson Jr, H.A. 
(eds.) Interstellar Processes.  Reidel, Dordrecht, p. 87\\
Launay J.M., Roueff E., 1977, JPhysB 10, 879\\
Lester D.F., Dinnerstein H.L., Werner M.W., et al., 1987, ApJ 320, 573\\
Lord S.D., Malhotra S., Lim T.L., et al., 1996, A\&A 315, L117\\
Madden S.C., Geis N., Genzel R., et al, 1993, ApJ 407, 579\\
Madden S., Geis N., Townes C.H.et al., 1995, Airbourne Astronomy
Symposium on the Galactic Ecosystem, ASP Conf. Series 73, 181. \\
Meyer D.M., Jura M., Cardelli J.A., 1998, ApJ 493, 222\\
Petuchowski S.J.,  Bennett C.L., 1993, ApJ 405, 591\\
Rubin R.H., 1985, ApJS 57, 349\\
Rubin R.H., Simpson J.P., Erickson E.F., et al., 1988, ApJ 327, 377\\
Rubin R.H., Simpson J.P., Lord S.D., et al., 1994, ApJ 420, 772 \\
Sofia U.J., Cardelli J.A., Guerin K.P., et al., 1997, ApJ 482, L105\\
Stacey G.J., Swain M.R., Bradford C.M., et al., 1999, 
The Universe as seen by ISO, ESA SP-427 p973\\
Swinyard B.M., Burgdorf M.J., Clegg P.E et al., 1998, SPIE 3354 \\
Tubbs A.D., 1980, ApJ 241, 969\\
van Gorkom J.H., van der Hulst J.M., Haschick A.D., 
et al., 1990 AJ 99, 1781\\
Wild W., Eckart A., Wilkind T., 1997, A\&A 322, 419\\
Wolfire M.G., Tielens A.G.G.M., Hollenbach D., 1990, ApJ 358,
116\\
Wolfire M.G., Tielens A.G.G.M., Hollenbach D., 1995, ApJ 443, 152\\


\end{document}